\documentclass[letterpaper,onecolumn,draftcls]{IEEEtran}

\usepackage[numbers,sort&compress]{natbib}

\makeatletter

\usepackage{etex}

\usepackage{zref-savepos}

\newcounter{mnote}

\def\xmarginnote{%
  \xymarginnote{\hskip -\marginparsep \hskip -\marginparwidth}}

\def\ymarginnote{%
  \xymarginnote{\hskip\columnwidth \hskip\marginparsep}}

\long\def\xymarginnote#1#2{%
\vadjust{#1%
\smash{\hbox{{%
        \hsize\marginparwidth
        \@parboxrestore
        \@marginparreset
\footnotesize #2}}}}}

\def\mnoteson{%
\gdef\mnote##1{\refstepcounter{mnote}\label{##1}%
  \zsavepos{##1}%
  \ifnum20432158>\number\zposx{##1}%
  \xmarginnote{{\color{blue}\bf $\langle$\arabic{mnote}$\rangle$}}%
  \else
  \ymarginnote{{\color{blue}\bf $\langle$\arabic{mnote}$\rangle$}}%
  \fi%
}
  }
\gdef\mnotesoff{\gdef\mnote##1{}}
\mnoteson
\mnotesoff

\usepackage{framed}
\usepackage{subcaption}
\usepackage{comment}
\usepackage[svgnames,dvipsnames]{xcolor}
\usepackage[all]{xy}
\usepackage{tikz}
\usetikzlibrary{positioning,matrix,through,calc,arrows,fit,shapes,decorations.pathreplacing,decorations.markings,}

\tikzstyle{block} = [draw,fill=blue!20,minimum size=2em]

\usepackage{qsymbols,amssymb,mathrsfs}
\usepackage{amsmath}
\usepackage[standard,thmmarks]{ntheorem}
\theoremstyle{plain}
\theoremsymbol{\ensuremath{_\vartriangleleft}}
\theorembodyfont{\itshape}
\theoremheaderfont{\normalfont\bfseries}
\theoremseparator{}

\theoremstyle{nonumberplain}
\theoremheaderfont{\scshape}
\theorembodyfont{\normalfont}
\theoremsymbol{\ensuremath{_\blacktriangleleft}}

\theoremnumbering{arabic}
\theoremstyle{plain}
\usepackage{latexsym}
\theoremsymbol{\ensuremath{_\Box}}
\theorembodyfont{\itshape}
\theoremheaderfont{\normalfont\bfseries}
\theoremseparator{}

\theorembodyfont{\upshape}
\theoremprework{\bigskip\hrule}
\theorempostwork{\hrule\bigskip}

\usepackage[overload]{empheq} 

\let\iftwocolumn\if@twocolumn
\g@addto@macro\@twocolumntrue{\let\iftwocolumn\if@twocolumn}
\g@addto@macro\@twocolumnfalse{\let\iftwocolumn\if@twocolumn}

\mathtoolsset{showonlyrefs=false,showmanualtags}
\let\underbrace\LaTeXunderbrace 
\let\overbrace\LaTeXoverbrace
\renewcommand{\eqref}[1]{\textup{(\refeq{#1})}} 
\newtagform{brackets}[]{(}{)}   
\usetagform{brackets}

\usepackage[Smaller]{cancel}

\PassOptionsToPackage{breaklinks,hyperindex=true,backref=false,bookmarksnumbered,bookmarksopen,linktocpage,colorlinks,linkcolor=BrickRed,citecolor=OliveGreen,urlcolor=Blue,pdfstartview=FitH}{hyperref}
\usepackage{hyperref}

\usepackage{graphicx,psfrag}
\graphicspath{{figure/}{image/}} 

\usepackage{diagbox} 

\usepackage{algorithm2e}
\usepackage{listings} 
\lstdefinelanguage{Maple}{
  morekeywords={proc,module,end, for,from,to,by,while,in,do,od
    ,if,elif,else,then,fi ,use,try,catch,finally}, sensitive,
  morecomment=[l]\#,
  morestring=[b]",morestring=[b]`}[keywords,comments,strings]
\DeclareMathAlphabet{\mathpzc}{OT1}{pzc}{m}{it}
\usepackage{upgreek} 
\usepackage{dsfont}  

\def\multi@nostar#1#2{%
  \expandafter\def\csname multi#1\endcsname##1{%
    \if ##1.\let\next=\relax \else
    \def\next{\csname multi#1\endcsname}     
    \expandafter\newcommand\csname #1##1\endcsname{#2}
    \fi\next}}

\def\multi@star#1#2{%
  \expandafter\def\csname #1\endcsname##1{#2}
  \multi@nostar{#1}{#2}
}

\newcommand{\multi}{%
  \@ifstar \multi@star \multi@nostar}

\multi*{rm}{\mathrm{#1}}
\multi*{mc}{\mathcal{#1}}
\multi*{op}{\mathop {\operator@font #1}}
\multi*{ds}{\mathds{#1}}
\multi*{set}{\mathcal{#1}}
\multi*{rsfs}{\mathscr{#1}}
\multi*{pz}{\mathpzc{#1}}
\multi*{M}{\boldsymbol{#1}}
\multi*{R}{\mathsf{#1}}
\multi*{RM}{\M{\R{#1}}}
\multi*{bb}{\mathbb{#1}}
\multi*{td}{\tilde{#1}}
\multi*{tR}{\tilde{\mathsf{#1}}}
\multi*{trM}{\tilde{\M{\R{#1}}}}
\multi*{tset}{\tilde{\mathcal{#1}}}
\multi*{tM}{\tilde{\M{#1}}}
\multi*{baM}{\bar{\M{#1}}}
\multi*{ol}{\overline{#1}}

\multirm  ABCDEFGHIJKLMNOPQRSTUVWXYZabcdefghijklmnopqrstuvwxyz.
\multiol  ABCDEFGHIJKLMNOPQRSTUVWXYZabcdefghijklmnopqrstuvwxyz.
\multitR   ABCDEFGHIJKLMNOPQRSTUVWXYZabcdefghijklmnopqrstuvwxyz.
\multitd   ABCDEFGHIJKLMNOPQRSTUVWXYZabcdefghijklmnopqrstuvwxyz.
\multitset ABCDEFGHIJKLMNOPQRSTUVWXYZabcdefghijklmnopqrstuvwxyz.
\multitM   ABCDEFGHIJKLMNOPQRSTUVWXYZabcdefghijklmnopqrstuvwxyz.
\multibaM   ABCDEFGHIJKLMNOPQRSTUVWXYZabcdefghijklmnopqrstuvwxyz.
\multitrM   ABCDEFGHIJKLMNOPQRSTUVWXYZabcdefghijklmnopqrstuvwxyz.
\multimc   ABCDEFGHIJKLMNOPQRSTUVWXYZabcdefghijklmnopqrstuvwxyz.
\multiop   ABCDEFGHIJKLMNOPQRSTUVWXYZabcdefghijklmnopqrstuvwxyz.
\multids   ABCDEFGHIJKLMNOPQRSTUVWXYZabcdefghijklmnopqrstuvwxyz.
\multiset  ABCDEFGHIJKLMNOPQRSTUVWXYZabcdefghijklmnopqrstuvwxyz.
\multirsfs ABCDEFGHIJKLMNOPQRSTUVWXYZabcdefghijklmnopqrstuvwxyz.
\multipz   ABCDEFGHIJKLMNOPQRSTUVWXYZabcdefghijklmnopqrstuvwxyz.
\multiM    ABCDEFGHIJKLMNOPQRSTUVWXYZabcdefghijklmnopqrstuvwxyz.
\multiR    ABCDEFGHIJKL NO QR TUVWXYZabcd fghijklmnopqrstuvwxyz.
\multibb   ABCDEFGHIJKLMNOPQRSTUVWXYZabcdefghijklmnopqrstuvwxyz.
\multiRM   ABCDEFGHIJKLMNOPQRSTUVWXYZabcdefghijklmnopqrstuvwxyz.

\newcommand{\dotleq}{\buildrel \textstyle  .\over {\smash{\lower
      .2ex\hbox{\ensuremath\leqslant}}\vphantom{=}}}
\newcommand{\dotgeq}{\buildrel \textstyle  .\over {\smash{\lower
      .2ex\hbox{\ensuremath\geqslant}}\vphantom{=}}}

\DeclareMathOperator*{\argmax}{arg\,max}

\newcommand{\bM}{\begin{bmatrix}}
\newcommand{\eM}{\end{bmatrix}}
\newcommand{\bSM}{\left[\begin{smallmatrix}}
\newcommand{\eSM}{\end{smallmatrix}\right]}
\renewcommand*\env@matrix[1][*\c@MaxMatrixCols c]{%
  \hskip -\arraycolsep
  \let\@ifnextchar\new@ifnextchar
  \array{#1}}

\newqsymbol{`N}{\mathbb{N}}
\newqsymbol{`R}{\mathbb{R}}
\newqsymbol{`Z}{\mathbb{Z}}

\newqsymbol{`|}{\mid}
\newqsymbol{`8}{\infty}
\newqsymbol{`1}{\left}
\newqsymbol{`2}{\right}
\newqsymbol{`6}{\partial}
\newqsymbol{`0}{\emptyset}
\newqsymbol{`-}{\leftrightarrow}

\DeclarePairedDelimiter\abs{\lvert}{\rvert}

\DeclarePairedDelimiter\Set{\{}{\}}
\DeclarePairedDelimiter\Span{\langle}{\rangle} 
\newcommand{\imod}[1]{\allowbreak\mkern10mu({\operator@font mod}\,\,#1)}

\newcommand{\threecols}[3]{
\hbox to \textwidth{%
      \normalfont\rlap{\parbox[b]{\textwidth}{\raggedright#1\strut}}%
        \hss\parbox[b]{\textwidth}{\centering#2\strut}\hss
        \llap{\parbox[b]{\textwidth}{\raggedleft#3\strut}}%
    }
}

\newcommand{\reason}[2][\relax]{
  \ifthenelse{\equal{#1}{\relax}}{
    \left(\text{#2}\right)
  }{
    \left(\parbox{#1}{\raggedright #2}\right)
  }
}

\newcommand{\utag}[2]{\mathop{#2}\limits^{\text{(#1)}}}
\newcommand{\uref}[1]{(#1)}

\let\SavedDoubleVert\relax
\let\protect\relax
{\catcode`\|=\active
  \xdef\extendvert{\protect\expandafter\noexpand\csname extendvert \endcsname}
  \expandafter\gdef\csname extendvert \endcsname#1{\mskip-5mu \left.%
      \ifx\SavedDoubleVert\relax \let\SavedDoubleVert\|\fi
     \:{\let\|\SetDoubleVert
       \mathcode`\|32768\let|\SetVert
     #1}\:\right.\mskip-5mu}
}
\def\SetVert{\@ifnextchar|{\|\@gobble}
    {\egroup\;\mid@vertical\;\bgroup}}
\def\SetDoubleVert{\egroup\;\mid@dblvertical\;\bgroup}

\begingroup
 \edef\@tempa{\meaning\middle}
 \edef\@tempb{\string\middle}
\expandafter \endgroup \ifx\@tempa\@tempb
 \def\mid@vertical{\middle|}
 \def\mid@dblvertical{\middle\SavedDoubleVert}
\else
 \def\mid@vertical{\mskip1mu\vrule\mskip1mu}
 \def\mid@dblvertical{\mskip1mu\vrule\mskip2.5mu\vrule\mskip1mu}
\fi

\makeatother

\usepackage{fouridx}
\usepackage{framed}
\usetikzlibrary{positioning,matrix}

\usepackage{paralist}
\usepackage{enumerate}

\usepackage[normalem]{ulem}

{\endMakeFramed}

\newenvironment{ybox}{
	\setlength{\FrameSep}{1.5mm}
	\setlength{\FrameRule}{0mm}
  \MakeFramed {\FrameRestore}}%
{\endMakeFramed}

\newenvironment{gbox}{
	\setlength{\FrameSep}{1.5mm}
\setlength{\FrameRule}{0mm}
  \MakeFramed {\FrameRestore}}%
{\endMakeFramed}

{\endMakeFramed}

 {\endMakeFramed}

\usepackage{enumitem}

\usepackage{etoolbox}

\let\theparentequation\theequation
\patchcmd{\theparentequation}{equation}{parentequation}{}{}

\renewenvironment{subequations}[1][]{
	\refstepcounter{equation}%
	\setcounter{parentequation}{\value{equation}}
	\setcounter{equation}{0}
	\def\theequation{\theparentequation\alph{equation}}%
	\let\parentlabel\label
	\ifx\\#1\\\relax\else\label{#1}\fi
	\ignorespaces
}{%
	\setcounter{equation}{\value{parentequation}}
	\ignorespacesafterend
}

\newcommand*{\nextParentEquation}[1][]{
	\refstepcounter{parentequation}
	\setcounter{equation}{0}
	\ifx\\#1\\\relax\else\parentlabel{#1}\fi
}

\newcommand{\CS}{C_{\op{S}}}

\title{Multiterminal Secret Key Agreement with Nearly No Discussion}

\author{Chung Chan, Manuj Mukherjee, Praneeth Kumar Vippathalla, and Qiaoqiao Zhou
  \thanks{Parts of this work were presented at the 2018 IEEE International Symposium on Information Theory (ISIT 2018), Vail, Colorado, U.S.A.~\cite{chan18isit}.}
	\thanks{Chung Chan (email: chung.chan@cityu.edu.hk) is with the Department of Computer Science, City University of Hong Kong.}
  \thanks{Manuj Mukherjee is with D\'{e}partement Communications \& Electronique (COMELEC), Telecom ParisTech, Paris, France.}
  \thanks{Praneeth Kumar Vippathalla is with the Department of Electrical Communication Engineering, Indian Institute of Science, Bangalore 560012, India.}
  \thanks{Qiaoqiao Zhou is with the Institute of Network Coding and the Department of Information Engineering, the Chinese University of Hong Kong.
  }
}

\begin{document}

\IEEEoverridecommandlockouts
\maketitle

\begin{abstract}
  We consider the secret key agreement problem under the multiterminal source model proposed by Csisz\'ar and Narayan. A single-letter characterization of the secrecy capacity is desired but remains unknown except in the extreme case with unlimited public discussion and without wiretapper's side information. Taking the problem to the opposite extreme by requiring the public discussion rate to be zero asymptotically, we obtain the desired characterization under surprisingly general setting with wiretapper's side information, silent users, trusted and untrusted helpers. An immediate consequence of the result is that the capacity with nearly no discussion is the same as the capacity with no discussion, resolving a previous conjecture in the affirmative. The idea of the proof is to characterize the capacity in the special case with neither wiretapper's side information nor untrusted helpers using a multivariate extension of G\'acs-K\"orner common information, and then extend the result to the general setting by a change of scenario that turns untrusted helpers into trusted helpers. We further show how to evaluate the capacity explicitly for finite linear sources and discuss how the current result can be extended to improve and unify existing bounds on the capacity for strictly positive discussion rates.
\end{abstract} 

\begin{IEEEkeywords}
  Multiterminal secret key agreement; G\'acs--K\"orner common information; secrecy capacity; double Markov inequality; general source model; finite linear source.
\end{IEEEkeywords}

\section{Introduction}
\label{sec:introduction}

The problem of secret key agreement between two users was formulated close to thirty years ago in \cite{maurer93,ahlswede93}, following a counter-intuitive result pointed out in \cite{bennett1988privacy} that two users can extract a longer secret key from their private observations by discussing in public. In particular, under the basic source model where two users observe i.i.d.\ sequences of the private correlated random sources, say $\RX$ and $\RY$, respectively, the maximum secret key rate, referred to as the secrecy capacity, is equal to the mutual information $I(\RX\wedge \RY)$ between the correlated sources. If the users cannot discuss in public, the problem reduces to the one considered in \cite{gacs72}, where the capacity is the entropy of the maximal common function of the correlated sources, referred to as the G\'acs--K\"orner common information $J_{\op{GK}}(\RX\wedge \RY)$. It should be mentioned that a more general setting considers a wiretapper with side information, say $\RZ$, in addition to listening to the entire public discussion, but there is hitherto no single-letter characterization of the capacity except in the special case when the discussion is one-way~\cite[Theorem~1]{ahlswede93}. Nevertheless, extensions of the model beyond the two-user case are of practical and theoretical interests because, as can been seen from the results in~\cite{chan15mi,chan17oo,chan18isit}, the problem calls for multivariate extensions of various well-known information measures in the bivariate case, including Shannon's mutual information, G\'acs--K\"orner common information and Wyner common information~\cite{wyner75}.

The simplest extension of secret key agreement beyond two users is by \cite{csiszar00}, which, in addition to the two active users who attempt to share a secret key, introduces a helper who can help discuss in public but need not share the secret key. The general multiterminal setting was subsequently formulated in \cite{csiszar04}, allowing arbitrary numbers of active users and helpers. There are also untrusted helpers whose observations may be leaked to the wiretapper. In the case without wiretapper's side information, and when the public discussion is interactive and unlimited in rate, the secrecy capacity has a single-letter characterization in \cite{csiszar04}. The capacity was studied and proposed in \cite{chan2008tightness,chan15mi} as an extension of Shannon's mutual information to the multivariate case, with application to data clustering in~\cite{chan16cluster}. A theoretically appealing characterization using the residual independence relation can be found in \cite[Theorem~5.1]{chan15mi}. The model in \cite{amin10a} considers a new scenario where a proper subset of active users are silent, i.e., not allowed to discuss in public, but the model does not have helpers. Similar to \cite{csiszar00}, the capacity was characterized~\cite{amin10a} in the case with unlimited public discussion and without wiretapper's side information. For the setting with helpers in addition to silent active users, the capacity is characterized in~\cite[Theorem~5.1]{chan17ooa} (the longer version of~\cite{chan17oo}). Once again, no characterization is known for the case with wiretapper's side information. Even the case with two users remains unsolved beyond one-way discussion.

Another challenge is to characterize the capacity as a function of the public discussion rates. For the two-user case with one-way discussion, the capacity was characterized in \cite[Theorem~2.6]{csiszar00}, extending the result in \cite{ahlswede93} for unlimited one-way discussion. A more explicit characterization was also derived for Gaussian sources in \cite{watanabe10,watanabe11}. When the discussion is asymptotically zero, the capacity reduces to the G\'ac--K\"orner common information in the two user case, which can be achieved with no discussion. The characterization also applies to the two-user case with interactive discussion as a consequence of the more general characterization of the capacity in \cite{LCV16} as a function of the public discussion rates. For the multiterminal case without helpers nor silent users, it was conjectured in \cite{chan18isit} that the capacity is equal to a multivariate extension of the G\'ac--K\"orner common information, which can be achieved again without any discussion. However, the conjecture remains open except for the special finite linear source models~\cite{chan18isit}. 

In this work, we aim to resolve the conjecture and characterize the secrecy capacity for the general source model with interactive discussion at asymptotically zero discussion rate in the presence of wiretapper's side information, helpers and silent users. The problem covers the case when no discussion is allowed, i.e., when all users are silent. To the best of our knowledge, no existing models or results directly cover the problem in such generality. In particular, the model in \cite{amin10a} and \cite{chan17ooa} do not cover the case where all users are silent as the proof technique relies on having at least one user with unbounded discussion rate. There is also no obvious multiterminal extension of the capacity characterizations in the case of two active users~\cite{ahlswede93,csiszar00,LCV16}, especially the converse proofs that rely on the Csisz\'ar-sum identity. Furthermore, the characterization~\cite{LCV16} for interactive discussion does not involve wiretapper's side information. The characterization in \cite{LCV16} is based on the idea of \cite{tyagi13} that uses the interactive source coding result of \cite{Kaspi}. The characterization is hard to evaluate as it involves a large number of auxiliary random variables that grows in the number of rounds of interactive discussion, which may go unbounded. There are other bounding techniques for the multiterminal secrecy capacity such as the lamination bound in \cite[Theorem~4.3]{chan17isit}\cite{chan19lamination} and the helper-set bound in \cite[Theorem~4.1]{chan17isit}. However, lamination does not extend beyond hypergraphical sources, while the helper-set bound was shown to be loose for a simple example in \cite[Fig.~2]{chan17isit} at asymptotically zero discussion rate. Despite all these challenges, we found that the capacity at asymptotically zero discussion has a simple characterization and resolved the previous conjecture in the affirmative as a special case. 

The paper is organized as follows: We will formulate the problem in Section~\ref{sec:problem}, and give the main results in Section~\ref{sec:results} followed by some discussion of the results in Section~\ref{sec:discussion} and their proofs in the Appendix.

\section{Problem formulation}
\label{sec:problem}

We shall consider the multiterminal source model for secret key generation introduced in \cite{csiszar04}, which is specified by a finite set $V$ of $\abs{V}\geq 2$ users and a discrete memoryless multiple source 
\begin{align*}
  \RZ_V:=(\RZ_i\mid i\in V)
\end{align*}
taking values in the finite alphabet $Z_V:=\prod_{i\in V}Z_i$, and distributed jointly according to $P_{\RZ_V}$. We remark here that we will be using the sans serif font for random variables and the normal font for their alphabet sets. The secret key agreement can be broken into a sequence of phases as follows. 

In the private observation phase, each user $i\in V$ observes $n$ i.i.d. samples $\RZ_i^n:=(\RZ_{i1},\dots,\RZ_{in})$ of the $i$th component source $\RZ_i$. In the private randomization phase, user~$i\in V$ can generate a private random variable $\RU_i$ independent of the private sources $\RZ_V^n$. Altogether 
\begin{align}
    P_{\RZ_V^n\RU_V}=P_{\RZ_V}^n\prod_{i\in V}P_{\RU_i}, \label{eq:src}
\end{align}
where $\RU_V=(\RU_i\mid i\in V)$.\footnote{The private randomization variables may be continuous.} To agree upon a secret key, all users except for a subset $S\subseteq V$ of silent users are allowed to communicate interactively over a public noiseless channel during the public communication phase. This implies that the communication sent by some user $i$ may depend on its accumulated observations. More precisely, at the $t$-th instant where $t\in \Set{1,2,\dots,r}$ for some chosen integer $r$, some vocal user $i_t\in V$ broadcasts a message $\tRF_t$ as a function of the previous messages $\tRF^{t-1}:=(\tRF_{\tau}\mid
\tau\leq t)$ and the private observation $(\RZ_{i_t}^n,\RU_{i_t})$ of the user $i_t$, i.e.,
\begin{align*}
  H(\tRF_t|\tRF^{t-1},\RZ_{i_t}^n,
  \RU_{i_t})=0.
\end{align*}
For convenience, we denote
the entire sequence of public messages by
\begin{align*}
  \RF &:= (\tRF_1,\tRF_2,\dots,\tRF_r).
\end{align*}
The rate of the public communication is given by $\limsup_{n\to\infty}\frac{1}{n}\log|F|$, where $F$ is the range of $\RF$.

Following the public communication, a predetermined subset $A\subseteq V$ of $\abs{A}\geq 2$ users  need to agree upon a secret key $\RK$ taking values in the set $K$. We also assume that another predetermined subset $D\subseteq V$, $D\cap A=\emptyset$, of users are being tapped by the wiretapper. The set $A$ is referred to as the set of \emph{active users}, the set $D$ is called the set of \emph{untrusted helpers}, whereas the users in $V\setminus (A\cup D)$ will be referred to as the \emph{trusted helpers}. We remark that each user in $A$ must be able to recover $\RK$ from its accumulated observations. On the other hand, any wiretapper listening to the public communication and having access to the untrusted helpers' observation, should be oblivious to $\RK$. In other words, we want $\RK$ to be `almost independent' of $(\RF,\RZ_D^n)$. More precisely, we need $\RK$ to satisfy the following recoverability and secrecy constraints: There exists some functions $`f_i$, for $i\in V$, such that
\begin{subequations}
\label{eq:recover-secrecy}
    \begin{align}
      \lim_{n\to`8} \Pr`1\{\exists i\in A, \RK\neq `f_i(\RF,\RZ_i^n,\RU_i)`2\}&=0 \label{eq:recover} & \text{(recoverability)}\\
      \limsup_{n\to `8} \frac1n`1[ \log \abs{K} - H(\RK|\RF,\RZ_D^n,\RU_D)`2] &= 0.\label{eq:secrecy} & \text{(secrecy)}
    \end{align}
\end{subequations}
The rate of the secret key $\RK$ is defined to be $\liminf_{n\to\infty}\frac{1}{n}\log|K|$. We define the secrecy capacity with a total communication rate $R\geq 0$ by
\begin{subequations}
\label{eq:CS}
    \begin{align}
      \CS(R) &:= \liminf_{n\to `8} \frac1n \log \abs {K} \quad
               \text{such that} \label{eq:CS:K}\\
      &\limsup_{n\to`8} \frac1n \log \abs{F} \leq R. \label{eq:R}
    \end{align}
\end{subequations}
We are interested in characterizing $\CS(0)$, namely, the secrecy capacity with asymptotically zero discussion rate. It is important to point out that our formulation covers the model with wiretapper's side information by the sources $\RZ_{S\cap D}$ of the silent untrusted helpers. This is because having a wiretapper observe some side information $\RZ$ directly is equivalent to having the wiretapper observe it through the source of a silent untrusted helper. Since a silent untrusted user cannot discuss, its knowledge of the side information cannot affect the secrecy capacity. The formulation also cover the case with no discussion by allowing $S$ to be the entire set $V$, unlike \cite{amin10a,chan17ooa} which require the $S$ to be a proper subset of $A$.

We should remark here that the secrecy constraint appearing in \eqref{eq:secrecy} is referred to as \emph{weak secrecy} in the literature. Several works including \cite{csiszar04} study a stronger secrecy criteria, referred to as \emph{strong secrecy}, obtained by removing the $\frac{1}{n}$ term from \eqref{eq:secrecy}. Our results are valid for both the weak secrecy and the strong secrecy criteria. We choose to define secrecy using the weak secrecy criteria since the main bottleneck in our proof is the converse part, i.e., obtaining an upper bound on $\CS(0)$. Noting that a key satisfying strong secrecy will by default satisfy weak secrecy, an upper bound on $\CS(0)$ defined using weak secrecy will therefore automatically translate to an upper bound on $\CS(0)$ defined using strong secrecy.

\section{Main results}
\label{sec:results}

The main result is the following single-letter characterization of the secrecy capacity at asymptotically zero discussion rate, in the presence of active users $A$, trusted helpers $(V`/A)`/D$, untrusted helpers $D$ and silent users $S$:
\begin{Theorem}
  \label{thm:CS0}
  The secrecy capacity at asymptotically zero discussion rate is
 \begin{align}
  \CS(0)=H(\RG|\RZ_D), \label{eq:CS0}
  \end{align}
  where $\RG$ is a solution to
  \begin{align}
      J_{\op{GK}}(\RZ_A):=\max\Set{H(\RG)\mid H(\RG|\RZ_i)=0,\forall i\in A}. \label{eq:JGK}
  \end{align}
  Furthermore, the capacity can be achieved with no discussion.
\end{Theorem}

Note that the capacity does not depend on $S$, i.e., the capacity remains unchanged whether a user is silent or not. This is consistent with the fact that the capacity can be achieved without discussion. Furthermore, notice that the capacity does not depend on the sources $\RZ_{(V`/A)`/D}$ of the trusted helpers because the solution $\RG$ to \eqref{eq:JGK} depends only on the sources $\RZ_A$ of the active users. In other words, the capacity remains unchanged even if the trusted helpers were removed, i.e., with $V$ reassigned as $A\cup D$. This is expected because, according to the secret key agreement protocol, helpers need not share the secret key but may help improve the secrecy capacity via public discussion. However, the fact that the capacity can be achieved with no discussion means that the trusted helpers cannot improve the capacity by discussion. Similarly, untrusted helpers cannot increase the capacity by discussion but their presence may diminish the capacity because the capacity $H(\RG|\RZ_D)$ with untrusted helpers is no larger than the capacity $ H(\RG)$ without untrusted helpers. This is again expected because the sources $\RZ_D$ of the untrusted helpers are leaked to the wiretapper, and so the common randomness between $\RG$ and $\RZ_D$ cannot be used for the secret key.

$J_{\op{GK}}(\RZ_A)$ in \eqref{eq:JGK} is a multivariate extension of the G\'acs-K\"orner common information first introduced by G\'acs and K\"orner in \cite{gacs72} for the case of two users. The optimal $\RG$ is unique up to bijections and
referred to as the \emph{maximal common function} (m.c.f.) of $\RZ_i$ for $i\in A$. The fact $\RG$ is called a common function is because the constraint $H(\RG|\RZ_i)=0$ in \eqref{eq:JGK} implies $\RG=g_i(\RZ_i)$ for some function $g_i$. It is called maximal because, if there exists another common function $\RG'$ that is not a function of $\RG$, i.e., $H(\RG'|\RG)>0$, then $H(\RG',\RG)>H(\RG)$, leading to the contradiction that $(\RG',\RG)$ is a strictly better solution to \eqref{eq:JGK}. Once again, the fact that the capacity can be achieved with no discussion is consistent with its characterization via the maximal common function that every active user can compute from their source without discussion. We remark that, while it is obvious the characterization of the capacity is achievable with no discussion, proving that the characterization is the best achievable rate is non-trivial, especially when public discussion, albeit of zero rate, is allowed. We will give the proof of the main result in Appendix~\ref{sec:proof:CS0}, and an alternative proof for the case with no discussion in Appendix~\ref{app:zero} to explain the non-triviality involved in handling the case with public discussion.

 While a single-letter characterization is widely accepted as a computable solution in Information Theory, the computation is often very difficult due to optimization over auxiliary random variables such as the maximal common function  $\RG$ in our case. For the characterization to be useful, it is important to be able to compute it efficiently. Fortunately, there does exist a systematic method called the \emph{ergodic decomposition} to compute the G\'ac--K\"orner common information \cite{gacs72}, and such a method can be directly extended to the multivariate case using an inductive argument, similar to the inductive proofs in the appendix. However, the computation is exponential in the number of random variables, and it is hard to give an explicit expression for the G\'ac--K\"orner common information for large networks. For the remainder of this section, we introduce a broad class of correlated random sources, called the \emph{finite linear source} model, and give a polynomial-time computable expression for the maximal common function and therefore the secrecy capacity.

\begin{Definition}[\mbox{\cite{chan10phd}}]
  \label{def:fls}
  A source $\RZ_V$ is said to be a finite linear source if its component can be written up to bijections\footnote{Two random variables $\RZ$ and $\RZ'$ are said to be bijections of each other iff $H(\RZ|\RZ')=H(\RZ'|\RZ)=0$.} as
  \begin{align}
    \RZ_i = \RMx \MM_i\label{eq:fls} \quad \forall i\in V,
  \end{align}
  where 
  $\RMx$ is a uniform random vector with elements taking
  values from some finite field $\bbF_q$, and $\MM_i$ is a
  deterministic matrix with elements from $\bbF_q$. 
\end{Definition}

\begin{Theorem}
  \label{thm:fls}
  For finite linear sources, the solution to \eqref{eq:JGK}, i.e., the m.c.f.\ of $\RZ_i$ for $i\in A$, is given by
  \begin{equation}
    \RG = \RMx \MM,
    \label{eq:G:fls}
  \end{equation}
  where $\MM$ is a matrix whose column space is 
  \begin{align}
       \Span{\MM}=\bigcap_{i\in A} \Span{\MM_i},   \label{eq:M}
  \end{align}
  namely the intersection of the column spaces of all $\MM_i$ for $i\in A$. $\Span{\MM}$ is also the maximum common subspace equal to $\argmax_{S} \Set{\dim S \mid S\subseteq \Span{M_i} \,\forall i\in A}$.
  
  Therefore, the G\'ac--K\"orner common information is given by 
  \begin{equation}
      J_{\op{GK}}(\RZ_A)=\op{rank}(\MM)\log q, \label{eq:JGK:fls}
  \end{equation}
  namely the dimension of the maximum common subspace in $\log q$~bits.
\end{Theorem}
The proof of Theorem~\ref{thm:fls} is given in Appendix~\ref{sec:proof:fls}. In the presence of untrusted helpers, the secrecy capacity in \eqref{eq:CS0} is simply
\begin{align*}
    H(\RG|\RZ_D) &=H(\RG,\RZ_D)-H(\RZ_D)\\
    &= \log q `1[\op{rank}(\bSM \MM & \MM_D\eSM) - \op{rank}(\MM_D)`2],
\end{align*}
where $\MM$ is a matrix whose column space satisfies \eqref{eq:M}. We conclude this section by giving an example of a finite linear source and computing its G\'ac--K\"orner common information. 

\begin{Example}
  \label{eg:fls:2}
  Let $\RX_a$, $\RX_b$, and $\RX_c$ be uniformly random and
  independent bits. Consider $A=V=\Set{1,2}$ ($D=`0$), and set
  \begin{align*}
    \RZ_1 &:= (\RX_a, \RX_b, \RX_a\oplus \RX_b)\\
    \RZ_2 &:= (\RX_c,\RX_a\oplus \RX_b\oplus\RX_c).
  \end{align*}
  This is a finite linear source because, with $\RMx:=\bSM \RX_a,\RX_b,\RX_c\eSM$,
  \begin{align*}
    \RZ_1 =\RMx \overbrace{\bM 1 & 0 & 1\\ 0 & 1 & 1\\ 0 & 0 &
                                                               0\eM}^{\MM_1:=},\quad
    \RZ_2 = \RMx\overbrace{ \bM 0 & 1 \\ 0 & 1\\ 1 & 1 \eM}^{\MM_2:=},
  \end{align*}
  and $\RMx$ is uniformly distributed over $\bbF_2^2$. Note also that $\CS(0)=J_{\op{GK}}(\RZ_A)$ by \eqref{eq:CS0}.

  Before computing
  $\RG$ in \eqref{eq:G:fls}, notice that $\MM_1$ does not have full
  column rank because the last column is the sum of the first two. We
  may remove the last column and consider instead
  \begin{align}
    \RZ_1 &= \RMx \overbrace{\bM 1 & 0 \\ 0 & 1 \\ 0 & 0
                                                       \eM}^{\MM_1:=}\text{
                                                       and }
                                                       \RZ_2 = \RMx\overbrace{ \bM 0 & 1 \\ 0 & 1\\ 1 & 1 \eM}^{\MM_2:=}.\notag
  \end{align}
  To compute
  $\Span{\MM_1}\cap \Span{\MM_2}$, note that the null space of
  $\bSM \MM_1 &
  \MM_2 \eSM= `1[`1.\begin{smallmatrix} 1 & 0 \\ 0 & 1 \\ 0 & 0 \end{smallmatrix}`2|\begin{smallmatrix} 0 & 1\\ 0 & 1\\ 1 & 1\end{smallmatrix}`2]$ is
  spanned by $\bM \Mu \\ \Mv \eM $ with $\Mu=\Mv=\bSM 1 \\ 1
  \eSM$. Therefore, the matrix
                                      \begin{align}
                                        \MM:=\MM_1\Mu = -\MM_2\Mv = \bM1 & 1 & 0\eM^\intercal         \label{eq:fls:2:M}                               
                                      \end{align} 
spans the desired intersection $\Span{\MM_1}\cap \Span{\MM_2}$. Hence, $\RG
= \RMx\MM = \RX_a\oplus \RX_b$. Hence, by \eqref{eq:JGK:fls}, we have $J_{\op{GK}}(\RZ_A)=1$.
  \end{Example}

  \section{Discussion}
  \label{sec:discussion}
  
  
  The primary goal of multiterminal secret key agreement is to understand how users should discuss to share a secret key not known to a wiretapper. By characterizing the secrecy capacity as a function of the public discussion rates of individual users, we gained valuable insights of the theoretical limits and the achieving schemes. The characterization of upper and lower bounds also inspired meaningful information measures and their properties applicable to other related problems. 
  Despite the challenges of characterizing the capacity in the two-user case under the basic source model, we obtained a simple and meaningful characterization by requiring the discussion rate to go to zero asymptotically. The characterization is a result of a better understanding of the G\'acs--K\"orner common information and its appropriate multivariate extension. 
  
  In contrast to the result~\cite{bennett1988privacy} that public discussion improves the secret key rate, our work conveys the opposite message that one cannot improve the secret key rate by public discussion at asymptotically zero discussion rate. Despite such a negative result, our work demonstrates how one can characterize the secrecy capacity in the multiterminal case with public discussion at limited rate. While this work focuses on asymptotically zero discussion rate, the proof techniques can be extended to the case with strictly positive discussion rate. There are various existing bounds on the secret capacity for positive discussion rate but they have obvious limitations. For the multiterminal setting, the best upper bounds are the helper-set bound and lamination bounds in \cite{chan17isit,chan19lamination}. While the helper-set bound is tight for a special class of pairwise independent networks (PIN)~\cite[Theorem~4.2]{chan17isit}, it is loose for a simple PIN~\cite[Fig.~2]{chan17isit} at zero discussion rate. Our result is expected to improve the helper-set bound strictly because it characterizes the capacity at zero discussion rate. Even though the basic lamination bound~\cite[Theorem~4.3]{chan17isit} is already tight for the general PIN model~\cite[Theorem~4.4]{chan17isit}, the lamination bound makes use of seemingly different techniques~\cite[Lemma~A.1]{chan19lamination} that do not apply beyond hypergraphical sources, even to finite linear sources. Our result is expected to improve the helper-set bound for general sources. It may potentially lead to a unified bound that covers the lamination bound for hypergraphical sources and also the lower bound for communication complexity in \cite{chan17oo} via the Wyner common information. 
  
  It difficult to extend our result to give a single-letter characterization for positive discussion rates as the two-user case remains unsolved~\cite{LCV16}. However, we believe it is possible to resolve the conjecture in \cite{mukherjee16} that the decremental secret key agreement scheme in \cite{chan16isit} achieves the capacity for hypergraphical sources. In particular, the resulting characterization of the communication complexity may be viewed as an asymptotic counter-part of that of \cite{courtade16} for non-asymptotic hypergraphical sources, but without the assumption that the discussion is linear.  Note that the decremental secret key agreement scheme can be extended to the compressed secret key agreement scheme in \cite{chan17cska}. Therefore, a more general result applicable beyond hypergraphical sources would be an optimality condition for compressed secret key agreement that can be satisfied for any hypergraphical sources using decremental secret key agreement. We remark that the capacity was characterized for the PIN model~\cite[Theorem~4.4]{chan17isit} only in the case without helpers, as the achieving scheme uses the tree-packing protocol~\cite{nitinawarat10}, which is not optimal in the case with helpers. Hence, a less ambitious goal is to characterize the capacity for the PIN model with helpers. 
  

\appendices

\numberwithin{equation}{section}
\makeatletter
\@addtoreset{equation}{section}
\renewcommand{\theequation}{\thesection\arabic{equation}}
\@addtoreset{Theorem}{section}
\renewcommand{\theTheorem}{\thesection\arabic{Theorem}}
\@addtoreset{Lemma}{section}
\renewcommand{\theLemma}{\thesection\arabic{Lemma}}
\@addtoreset{Corollary}{section}
\renewcommand{\theCorollary}{\thesection\arabic{Corollary}}
\@addtoreset{Proposition}{section}
\renewcommand{\theProposition}{\thesection\arabic{Proposition}}
\makeatother

\section{Proof of Theorem~\ref{thm:CS0}}
\label{sec:proof:CS0}

In this section, we derive the characterization~\eqref{eq:CS0} of the secrecy capacity at asymptotically zero discussion rate and show that the capacity can be achieved with neither discussion nor private randomization. We first show the achievability, i.e., one can choose the secret key at rate $H(\RG|\RZ_D)$ with no public discussion while satisfying the recoverability and secrecy constraints in~\eqref{eq:recover-secrecy}. By the balanced coloring Lemma \cite[Lemma~B.3]{csiszar04}\footnote{We set $\RU$, $\RV$, and $g$ in \cite[Lemma~B.3]{csiszar04} to $\RG$, $\RZ_D$ and a constant function respectively.}, there exists a choice of $\RK$ satisfying the conditions
\begin{align*}
    \lim_{n\to `8} \log \abs{K} - H(\RK|\RZ_D^n) = 0\\
    \lim_{n\to `8} \frac1n\log \abs{K} - H(\RG|\RZ_D) = 0.
\end{align*}
The first equality implies the secrecy constraint~\eqref{eq:secrecy} while the recoverability constraint~\eqref{eq:recover} follows from the fact that $\RG$ is a common function computable by the active users with no discussion. The last equality implies that the secret key rate is $H(\RG|\RZ_D)$ as desired. This completes the proof of achievability.

For the converse proof, it suffices to consider the case without silent users, i.e., $S=`0$, because the bound for this case also applies to the case with silent users. Compared to the proof of achievability, the converse proof is more complicated and will be broken into two steps: We first prove for the case without untrusted helpers that $\CS(0)\leq J_{\op{GK}}(\RZ_A)$;
Then, we consider the case with untrusted users and extend the result to prove $\CS(0)\leq H(\RG|\RZ_D)$ in \eqref{eq:CS0}. More precisely, to make use of the result for the case without untrusted helpers, we will consider a change of scenario that turns untrusted helpers into trusted helpers. Roughly speaking, if $\RK$ is a feasible secret key for the original scenario with untrusted helpers, it is also a feasible secret key for the modified scenario. Since the key rate for the modified scenario with no untrusted helper is upper bounded by G\'ac--K\"orner common information, we can argue that $\frac1n H(\RK|\RG^n)$ goes to $0$, i.e., the randomness in $\RK$ is primarily from that of the m.c.f.\ $\RG$. This will imply the desired capacity upper bound $H(\RG|\RZ_D)$ for the original scenario with untrusted helpers because the randomness in $\RZ_D$ cannot be used for the secret key.

 We remark that our approach is different from the converse proof in \cite{csiszar00} that handles the wiretapper's side information, or equivalently, the source of silent untrusted helpers by the Csisz\'ar-sum identity. It appears that the technique using Csisz\'ar-sum identity is limiting and does not extend to the multiterminal setting involving more than two active users.

\subsection{Proof of converse without untrusted helpers}
\label{sec:proof:CS0:trusted}

The converse proof for the case with untrusted helpers follows a similar single-letterization technique as in \cite{csiszar00} and uses the following property of the m.c.f.: 
\begin{Lemma}
\label{lem:mar}
For $A\subseteq B\subseteq V$, the Markov conditions
\begin{align}
I(\RQ\wedge \RZ_B|\RZ_i)=0 \quad \forall i \in A
\label{eq:mmc}
\end{align}
imply the Markov condition
\begin{align}
I(\RQ\wedge \RZ_B|\RG)=0,\label{eq:mc}
\end{align}
where $\RG$ is the m.c.f.\ of $\RZ_i$ for $i\in A$.
\end{Lemma}
The Lemma can be viewed as a multivariate extension of the double Markov inequality~\cite[Problem 16.25]{csiszar2011information}, which is the special case when  $\abs{A}=2$ and $A=V$. 
\begin{Proof}
The Markov conditions \eqref{eq:mmc} means that $\RQ$ and $\RZ_B$ are independent given $\RZ_i$ for all $i\in A$, i.e.,
\begin{align*}
\underbrace{P_{\RQ|\RZ_i}(q|z_i)}_{f_i^{(q)}(z_i):=}=\underbrace{P_{\RQ|\RZ_B}(q|z_B)}_{f^{(q)}(z_B):=}\quad \forall q\in Q, z_B\in Z_B:P_{\RZ_B}(z_B)>0.
\end{align*}
It follows that 
\begin{align*}
f_i^{(q)}(\RZ_i)=f^{(q)}(\RZ_B)\quad \forall i\in A, q\in Q.
\end{align*}
Note that $f_i^{(q)}(\RZ_i)$ on the left is possibly random because $\RZ_i$ is. The above condition means that $f^{(q)}(\RZ_B)$ is a common function of $\RZ_i$ for $i\in A$, and so it must be a function of the maximal common function $\RG$ by \cite{gacs72}. (See also the explanation below \eqref{eq:JGK}.) Hence, 
\begin{align*}
f^{(q)}(\RZ_B) = P_{\RQ|\RG,\RZ_B}(q|\RG,\RZ_B)=P_{\RQ|\RG}(q|\RG) \quad \forall q\in Q,
\end{align*}
which implies the desired Markov condition~\eqref{eq:mc}. The first equality is because $\RG$ is a function of $\RZ_B$. The last equality is because $f_i^{(q)}(\RZ_i)$ is a function of $\RG$.
\end{Proof}
For the desired converse proof, we will apply the above Lemma with $A=B$. More precisely, we will show that
\begin{align}
    \CS(0) \leq \sup \Set{I(\RQ\wedge\RZ_A) \mid I(\RQ\wedge\RZ_A)=I(\RQ\wedge\RZ_i), \forall i\in A}, \label{eq:JGK:alt}
\end{align}
where the maximization is over all possible $P_{\RQ|\RZ_A}$.
Applying Lemma~\ref{lem:mar} with $B=A$, the constraint in \eqref{eq:JGK:alt} implies \eqref{eq:mmc} and so $I(\RQ\wedge \RZ_A|\RG)=0$ by \eqref{eq:mc}. By the data processing inequality,
\begin{align*}
    I(\RQ\wedge \RZ_A) \leq I(\RG\wedge \RZ_A) = H(\RG)
\end{align*}
and so \eqref{eq:JGK:alt} implies the desired bound $\CS(0)\leq H(\RG)=J_{\op{GK}}(\RZ_A)$.

It remains to prove \eqref{eq:JGK:alt}. Let $\RJ$ be uniformly distributed over $\{1,2,\ldots,n\}$ and independent of everything else, namely $(\RZ_V^n, \RU_V, \RK,\RF)$. Define
\begin{align*}
\RQ_{\RJ}:=(\RK,\RF,\RU_A, \RZ_{A}^{\RJ-1},\RJ).
\end{align*}

By the secrecy constraint~\eqref{eq:secrecy}, we have
\begin{align*}
\log\abs{K}&\leq H(\RK|\RF)+n\delta_n\\
		&= H(\RK,\RF)-H(\RF)+n\delta_n\\
		&=\underbrace{I(\RK,\RF\wedge \RZ_A^n)}_{`(1)}+\underbrace{H(\RK,\RF|\RZ_{A}^n)-H(\RF)}_{`(2)}+n\delta_n,
\end{align*}
where we use $\delta_n$ to denote a sufficiently large non-negative real number going to $0$ sufficiently slow as $n\to `8$. Next, we will bound $`(1)$ and $`(2)$ as follows:
On  one hand,
\begin{align*}  
	  `(1)&\leq I(\RK,\RF,\RU_A\wedge \RZ_A^n)\\
	  &=\sum_{j=1}^{n}I(\RK,\RF,\RU_A\wedge \RZ_{Aj}|\RZ_{A}^{j-1})\\
	  &=\sum_{j=1}^{n}I(\RK,\RF,\RU_A,\RZ_{A}^{j-1}\wedge \RZ_{Aj})\\
	  &=nI(\RQ_{\RJ}\wedge \RZ_{A\RJ})
\end{align*} 
where the second equality is because $I(\RZ_{A}^{j-1}\wedge \RZ_{A j})=0$, for all $j\in n$, by the memorylessness~\eqref{eq:src} of the source.
On the other hand, assuming $1,2\in A$ without the loss of generality,
\begin{align*}
`(2)&\utag{a}\leq H(\RK,\RF|\RZ_{A}^n,\RU_1)+H(\RK,\RF|\RZ_{A}^n,\RU_2)-H(\RF)\\
     &\utag{b}\leq H(\RF|\RZ_{A}^n,\RU_1)+ H(\RF|\RZ_{A}^n,\RU_2)-H(\RF)+n\delta_n\\
     &\utag{c}\leq H(\RF)+n\delta_n\\
     &\utag{d}\leq 2n\delta_n
\end{align*}
\uref{a} follows immmediately from the lamination bound in \cite[Lemma~A.1]{chan19lamination} because $\RU_1$, $\RU_2$, and $\RZ_A^n$ are mutually independent by \eqref{eq:src}. A more detailed derivation is as follows:
\begin{align*}
H(\RU_1|\RK,\RF,\RZ_A^n) &\geq H(\RU_1|\RK,\RF,\RZ_A^n,\RU_2)\quad \text{and so}\\
H(\RK,\RF,\RZ_A^n,\RU_1)+H(\RK,\RF,\RZ_A^n,\RU_2) &\geq H(\RK,\RF,\RZ_A^n)+H(\RK,\RF,\RZ_A^n,\RU_1,\RU_2)\\
&\geq H(\RK,\RF,\RZ_A^n)+H(\RZ_A^n,\RU_1,\RU_2)\\
&=H(\RK,\RF,\RZ_A^n)+H(\RZ_A^n,\RU_1)+H(\RZ_A^n,\RU_2)-H(\RZ_A^n),
\end{align*}
which, after rearrangement, leads to \uref{a}; 
\uref{b} follows from the fact that $$H(\RK|\RZ_{A}^n,\RU_1,\RF)+H(\RK|\RZ_{A}^n,\RU_2,\RF)\leq n\delta_n$$ by the recoverability constraint~\eqref{eq:recover}; \uref{c} is because conditioning cannot increase entropy, and \uref{d} is by the assumption that the discussion rate is asymptotically zero, i.e., $H(\RF)\leq \log{\abs{F}}\leq n\delta_n$. Therefore, 
\begin{align}
\label{eq:cs:q}
\frac{1}{n}\log \abs{K}\leq I(\RQ_{\RJ}\wedge \RZ_{A\RJ})+3\delta_n.
\end{align}

Since the discussion rate is asymptotically zero, we have for any $i\in A$
\begin{align*}
n\delta_n \geq \log\abs{F} &\geq H(\RF|\RZ_i^n,\RU_A)\\
		&\geq H(\RK,\RF|\RZ_i^n,\RU_A)-n\delta_n\\
	      &\geq I(\RK,\RF\wedge\RZ_A^n|\RZ_i^n,\RU_A)-n\delta_n\\
	      &=I(\RK,\RF\wedge\RZ_A^n|\RU_A)-I(\RK,\RF\wedge\RZ_i^n|\RU_A)-n\delta_n\\
	      &=I(\RK,\RF,\RU_A\wedge\RZ_A^n)-I(\RK,\RF,\RU_A\wedge\RZ_i^n)-n\delta_n\\
	      &=\sum_{j=1}^{n}I(\RK,\RF,\RU_A,\RZ_{A}^{j-1}\wedge \RZ_{Aj})-\sum_{j=1}^{n}I(\RK,\RF,\RU_A,\RZ_{i}^{j-1}\wedge \RZ_{ij})-n\delta_n\\
	      &\geq\sum_{j=1}^{n}I(\RK,\RF,\RU_A,\RZ_{A}^{j-1}\wedge \RZ_{Aj})-\sum_{j=1}^{n}I(\RK,\RF,\RU_A,\RZ_{A}^{j-1}\wedge \RZ_{ij})-n\delta_n\\
	      &=nI(\RQ_{\RJ}\wedge \RZ_{A\RJ})-nI(\RQ_{\RJ}\wedge \RZ_{i\RJ})-n\delta_n,
\end{align*}
where the third inequality follows from the recoverability constraint~\eqref{eq:recover}; the second equality is because $I(\RU_A\wedge\RZ_A^n)=I(\RU_A\wedge\RZ_i^n)=0$ by the assumption~\eqref{eq:src} of the private randomization; 
the third equality follows from the chain rule expansion and by the memorylessness~\eqref{eq:src} of the source.
Therefore,
\begin{align}
\label{eq:cs:makov}
I(\RQ_{\RJ}\wedge \RZ_{A\RJ})-I(\RQ_{\RJ}\wedge \RZ_{i\RJ})\leq \delta_n\quad \forall i\in A.
\end{align}

Combining~\eqref{eq:cs:q} and~\eqref{eq:cs:makov}, and using the fact that $P_{\RZ_{V\RJ}}=P_{\RZ_V}$, we have 
\begin{align*}
\CS(0) &\leq \limsup_{n\to `8} `G(`d_n) \quad \text{where}\\
`G(`d) &:= \sup \Set{I(\RQ\wedge\RZ_A)+\delta \mid I(\RQ\wedge\RZ_A)-I(\RQ\wedge\RZ_i)\leq \delta, \forall i\in A} \quad \text{for }`d\geq 0.
\end{align*}
The above maximization is over all possible choices of $P_{\RQ|\RZ_A}$. The solution exists because, using the Carath\'{e}odory-Fenchel-Eggleston Theorem, it can be argued that the support of $\RQ$ can be bounded uniformly for all $\delta>0$. (See for example \cite[Lemmas~3.4, 3.5]{csiszar2011information}.) $`G(`d)$ is also continuous in $`d$ by the continuity of the entropy function~\cite{csiszar81} for discrete random variables with finite alphabet sets. Hence,
\begin{align*}
    \limsup_{n\to `8} `G(`d_n)=`G(0) = \max \Set{I(\RQ\wedge\RZ_A) \mid I(\RQ\wedge\RZ_A)=I(\RQ\wedge\RZ_i), \forall i\in A}
\end{align*}
which gives \eqref{eq:JGK:alt} as desired. 

We remark that the above derivation does not invoke the Csisz\'ar-sum identity. The auxiliary random variable $\RQ$ comes from $\RQ_{\RJ}$, which is obtained by simple chain-rule expansion of the i.i.d.\ samples of the sources. We found that the problem of extending the single-letterization in \cite{csiszar00} using Csisz\'ar-sum identity is that the Csisz\'ar-sum identity involves expanding the sources in only two directions, which does not allow for a common definition of auxiliary random variable in the case with more than two active users.

\subsection{Proof of converse with untrusted helpers}
\label{sec:proof:CS0:untrusted}

In this section, we extend the above converse proof for the case without untrusted helpers to the case with untrusted helpers, i.e., $D\neq `0$. The proof relies on a technical Lemma \cite[Lemma~3.1]{csiszar08} similar to the proof of a rather different result~\cite[Theorem~3.2]{csiszar08} that the secret key can be chosen purely as a function of the source of any single active user. 

Consider any feasible secret key $\RK$ and discussion $\RF$ at asymptotically zero rate satisfying the recoverability and secrecy constraints~\eqref{eq:recover-secrecy}. Furthermore, assume $1\in A$ without loss of generality and let $\hat{\RK}:=`f_1(\RF,\RZ_1^n,\RU_1)$ be the secret key estimate generated by user~$1$. It follows from the recoverability constraint~\eqref{eq:recover} and Fano's inequality that
\begin{align}
    H(\hat{\RK}|\RF,\RZ_i^n,\RU_i) &\leq n`d_n \quad \forall i\in A \label{eq:K|F}\\
    H(\RK|\hat{\RK}) \leq n `d_n.\label{eq:K|hatK}
\end{align}
Again, we use $`d_n$ to denote a non-negative real number that is sufficiently large and that goes to $0$ sufficiently slowly as $n$ goes to infinity. 

Next, we modify the scenario by setting $D=`0$. $A$, $V$, $S$, and $\RZ_V$ remain unchanged. Instead of using $n$ to denote the block length, we will use $nn'$ as the block length where $n'$ is a positive integer. To distinguish the modified scenario from the original scenario, we will denote the secrecy capacity of the modified scenario by $\CS'$ instead of $\CS$. Similarly, we will use $\RK'$ and $\RF'$ to denote the secret key and public discussion for the modified scenario. By the converse proof in the previous section, we have \begin{align}
    \CS'(0)\leq J_{\op{GK}}(\RZ_A).\label{eq:CS'}
\end{align} 
We will show using the above bound that 
\begin{align}
    \lim_{n\to `8} \frac1n H(\RK|\RG^n) = 0  \label{eq:K|G}
\end{align}
and so, by the secrecy constraint~\eqref{eq:secrecy},
\begin{align*}
\frac{1}{n}\log|K|&\leq \frac{1}{n}H(\RK|\RF,\RZ_D^n,\RU_D)+\delta_n \\		
	 &\leq \frac{1}{n}H(\RK,\RG^n|\RZ_D^n)+\delta_n\\
	 &\leq \frac{1}{n}H(\RG^n|\RZ_D^n)+\frac{1}{n}H(\RK|\RG^n)+\delta_n\\
	 &\leq H(\RG|\RZ_D)+2\delta_n \quad\text{by \eqref{eq:K|G}},
\end{align*}
which implies $\CS(0)\leq H(\RG|\RZ_D)$, thereby establishing the desired result.

It remains to show \eqref{eq:K|G}, which means that the randomness in $\RK$ comes primarily from the m.c.f.\ $\RG$. Consider the modified scenario with $D=`0$. We first show that there exists a public discussion $\RF'$ at asymptotically zero rate such that $\hat{K}^{n'}$ can be recovered by every active user asymptotically in $n'$, i.e.,
\begin{align*}
    \Pr\Set{\exists i\in A, \hat{\RK}^{n'}\neq `f'_i(\RF',\RZ_i^{nn'},\RU_i^{n'})} \leq `d_{n'}
\end{align*}
for some decoding functions $`f'_A$.
In particular, we choose $\RF'=(\RF^{n'},\RF'')$ where $\RF$ is the discussion in the original scenario, and $\RF''$ is some additional discussion by user~$1$ at asymptotically zero rate. Existence of such $\RF''$ follows from \cite[Lemma~3.1]{csiszar08} by \eqref{eq:K|F}.\footnote{We set $\RU$, $\RV$, and $g$ in  \cite[Lemma~3.1]{csiszar08} to $\hat{\RK}$, $(\RF,\RZ_i^n,\RU_i)$, and $\RF''$ respectively.} 

Note that $(\hat{K}^{n'},\RG^{nn'})$ is also recoverable by every active user since $\RG$ is their common function. We can then extract a secret key $\RK'$ for the modified scenario from $(\hat{K}^{n'},\RG^{nn'})$ at rate $\frac1n H(\hat{K},\RG^n)$. More precisely, by the balanced coloring Lemma~\cite[Lemma~B.3]{csiszar04}, there exists a function $\RK'$ of $(\hat{K}^{n'},\RG^{nn'})$ satisfying
\begin{align*}
    \log \abs{K'}&\leq n'H(\RK'|\RF') + n'`d_{n'}\\
    \log\abs{K'} &\geq n'H(\hat{\RK},\RG^n|\RF) - n'`d_{n'},
\end{align*}
where the first inequality implies the desired secrecy constraint for the modified scenario. By the last inequality and the capacity bound~\eqref{eq:CS'} for the case without untrusted helpers, we have
\begin{align*}
    H(\RG^{nn'}) &\geq \log\abs{K'} - nn'`d_{nn'}\\
    &\geq n'H(\hat{\RK},\RG^n|\RF) - n'`d_{n'}-nn'`d_{nn'}
\end{align*}
We can further lower bound $H(\hat{\RK},\RG^n|\RF)$ by 
$H(\hat{\RK},\RG^n)-H(\RF)\geq H(\hat{\RK},\RG^n)-n`d_n$, where the last inequality is because $\RF$ has zero rate asymptotically in $n$. Rearranging the terms and letting $n'$ goes to infinity, we have
\begin{align*}
    H(\hat{K}|\RG^n) \leq n`d_n.
\end{align*}
By \eqref{eq:K|hatK}, $H(\RK|\RG^n)\leq 2n`d_n$, which implies \eqref{eq:K|G} as desired.

\section{Proof of Theorem~\ref{thm:fls}}
\label{sec:proof:fls}

To prove Theorem~\ref{thm:fls}, we shall make use of the following technical Lemma.

\begin{Lemma}
  \label{lem:fls}
  For any finite linear $(\RZ_1,\RZ_2)$, we have
  \begin{align}
      I(\RZ_1\wedge\RZ_2|\RMx\MM)=0  \label{eq:I|M}  
  \end{align}
  where $\MM$ is a matrix satisfying $\Span{\MM}=\Span{\MM_1}\cap\Span{\MM_2}$.
\end{Lemma}

\begin{Proof}
By standard arguments in linear algebra, there exists matrices $\MN_1$ and $\MN_2$ such that
\begin{align*}
    \Span*{\bM \MM & \MN_i \eM} &= \Span{\MM_i} \\
    \dim(\Span{\MM}\cap\Span{\MN_i}) &= 0
\end{align*}
for $i\in \Set{1,2}$. It follows that there is a bijection between $\RZ_i$ and $\RMx\bSM\MM & \MN_i \eSM$. To prove the Lemma, i.e., \eqref{eq:I|M}, it suffices to show that
\begin{align*}
    I(\RMx\MN_1\wedge\RMx\MN_2|\RMx\MM)=0.
\end{align*}
We will argue the stronger claim that $\RMx\MM$, $\RMx\MN_1$, and $\RMx\MN_2$ are mutually independent. Since $\RMx$ is uniformly random by Definition~\ref{def:fls}, it suffices to show
\begin{align*}
    \dim\biggl(\Span{\MM} \cap \Span{\MN_1}\biggr) = 0 = \dim\biggl(\Span{\bM \MM & \MN_1 \eM} \cap \Span{\MN_2}\biggr).
\end{align*}
The first equality implies $\RMx\MM$ is independent of $\RMx\MN_1$, while the second equality means that $(\RMx\MM,\RMx\MN_1)$ is independent of $\RMx\MN_2$ as desired. The first equality holds by the construction of $\MN_1$. The second equality holds because, otherwise, some column of $\MN_2$ is in $\Span{\bSM\MM &\MN_1\eSM}$ but not in $\Span{\MM}$, contradicting the fact that
\begin{align*}
    \Span{\bM \MM & \MN_1 \eM} \cap \Span{\bM \MM & \MN_2 \eM} = \Span{\MM_1} \cap \Span{\MM_2}=\Span{\MM}. 
\end{align*}
This completes the proof of the Lemma.
\end{Proof}

A interesting Corollary of the above Lemma is the following equivalence of bivariate G\'ac--K\"orner common information, Shannon mutual information, and Wyner common information.
\begin{Corollary}
\label{cor:fls}
    For any finite linear source $(\RZ_1,\RZ_2)$, we have 
    \begin{align}
        J_{\op{GK}}(\RZ_1\wedge \RZ_2) &= I(\RZ_1\wedge \RZ_2) = J_{\op{W}}(\RZ_1 \wedge \RZ_2)\label{eq:GIW}
    \end{align}
    where $J_{\op{GK}}$ and $J_{\op{W}}$ denotes the G\'ac--K\"orner common information and Wyner common information respectively:
    \begin{align}
        J_{\op{GK}}(\RZ_1\wedge \RZ_2)&:=J_{\op{GK}}(\RZ_{\Set{1,2}})=\max \Set{H(\RG)\mid H(\RG|\RZ_1)=H(\RG|\RZ_2)=0}\label{eq:G}\\
         J_{\op{W}}(\RZ_1\wedge \RZ_2) &:= \min \Set{I(\RW\wedge \RZ_1,\RZ_2)\mid I(\RZ_1\wedge \RZ_2|\RW)=0}.\label{eq:W}
    \end{align}
    Furthermore, the solution to \eqref{eq:G} and \eqref{eq:W} are given by $\RMx\MM$ in Lemma~\ref{lem:fls}.
\end{Corollary}

\begin{Proof}
 It was shown in \cite{wyner75} that \eqref{eq:GIW} holds with equalities replaced by $\leq$ in general for any sources. For the finite linear source, the reverse inequalities will follow by showing that $\RMx\MM$ in Lemma~\ref{lem:fls} is a solution to both \eqref{eq:G} and \eqref{eq:W} because that implies $H(\RMx\MM)\leq J_{\op{GK}}(\RZ_1\wedge \RZ_2)$ and $J_{\op{W}}(\RZ_1\wedge \RZ_2)\leq H(\RMx\MM)$.
 $\RMx\MM$ is a solution to \eqref{eq:G} because it is a common function of $\RZ_1$ and $\RZ_2$. $\RMx\MM$ is a solution to \eqref{eq:W} by \eqref{eq:I|M}.
\end{Proof}

Note that the above Corollary implies Theorem~\ref{thm:fls} for the case $\abs{A}=2$. We now prove Theorem~\ref{thm:fls} by induction on $|A|$. Assume the inductive hypothesis that for any $j\in A$, the m.c.f.\ of $\RZ_i$ for $i \in A`/\Set{j}$, i.e., the solution to $J_{\op{GK}}(\RZ_{A`/\Set{j}})$, is given by 
\begin{align*}
    \RMZ_j':=\RMx\MM^{(j)}\text{ where }\Span{\MM^{(j)}}=\bigcap_{i\in A`/\Set{j}}\Span{\MM_i}.
\end{align*}
It suffices to show that $H(\RG|\RMx\MM)=0$ as follows, since $\RMx\MM$ is a common function of $\RZ_i$ for $i\in A$ trivially:
\begin{align*}
    0 & \utag{a}= H(\RG|\RZ_j)+H(\RG|\RZ'_j)\\
      & \geq H(\RG|\RZ_j,\RMx\MM)+H(\RG|\RZ'_j,\RMx\MM)\\
      & = H(\RG,\RZ_j,\RMx\MM)+H(\RG,\RZ'_j,\RMx\MM)-H(\RZ_j,\RMx\MM)-H(\RZ'_j,\RMx\MM)\\
      & \utag{b}\geq H(\RG,\RZ_j,\RZ'_j,\RMx\MM)+H(\RG,\RMx\MM)-H(\RMx\MM)-H(\RZ_j|\RMx\MM)-H(\RZ'_j,\RMx\MM)\\
      & \utag{c}= H(\RG,\RZ_j,\RZ'_j,\RMx\MM)+H(\RG,\RMx\MM)-H(\RMx\MM)-H(\RMx\MM,\RZ_j,\RZ'_j)\\      
      & = H(\RG|\RZ_j,\RZ'_j,\RMx\MM)+H(\RG|\RMx\MM)\geq H(\RG|\RMx\MM).
\end{align*}
\uref{a} uses the fact that $\RG$ is a common function of $\RZ_j$ and $\RZ_j'$. \uref{b} is because $H(\RZ_j|\RG,\RZ'_j,\RMx\MM)\leq H(\RZ_j|\RG,\RMx\MM)$.  \uref{c} is by applying Lemma~\ref{lem:fls} with $\RZ_1$ and $\RZ_2$ in the Lemma set to $\RZ_j$ and $\RZ'_j$ respectively and noting that $\Span{\MM}=\Span{\MM^{(j)}}\cap \Span{\MM_j}$. This completes the proof.

We remark that Theorem~\ref{thm:fls} can be viewed as a partial extension of Corollary~\ref{cor:fls} to multiterminal finite linear sources. However, while \eqref{eq:GIW} with equalities replaced by $\leq$ continue to hold for the multivariate extensions of the mutual and common information measures~\cite[Corollary~6.2]{chan15mi}, the multivariate Wyner common information
\begin{align*}
    J_{\opW}(\RZ_V):=\min\Set*{\extendvert{I(\RW\wedge \RZ_V) | H(\RZ_V|\RW)=\sum_{i\in V} H(\RZ_i|\RW)}}
\end{align*}
can be strictly larger the multivariate mutual information
\begin{align*}
    I(\RZ_V):=\max\Set*{\extendvert{`g\in `R | H(\RZ_V)-`g \leq \sum_{C\in \mcP} [H(\RZ_C)-`g],\forall \mcP\in \Pi(V)}},
\end{align*}
where $\Pi(V)$ denotes the set of partitions of $V$.
A simple example can be constructed using the the characterization of the Wyner common in \cite{chan17oo} for the hypergraphical source model, which is a special case of the finite linear source model. Nevertheless, similar to the multivariate G\'ac--K\"orner common information, the multivariate Wyner common information can also be explicitly evaluated for finite linear sources using linear algebra.

\section{Alternative proof of Theorem~\ref{thm:CS0} for the case with no discussion}
\label{app:zero}

In this section, we extend the property \cite[Lemma~1.1]{csiszar00} of the m.c.f.\ to the multivariate case, which will lead to an alternative converse proof of Theorem~\ref{thm:CS0} in the case without private randomization nor public discussion is allowed, i.e., the case when $\RU_V$ is constant and $S=V$.

\begin{Lemma}
\label{lem:KG}
 Let $\RG$ be the m.c.f.\ of $\RZ_i$ for $i\in A$ with $\abs{A}\geq 2$, i.e., the solution to $J_{\op{GK}}(\RZ_A)$ in \eqref{eq:JGK}. For any $`e>0$, there exists $`d>0$ independent of $n$ such that
 \begin{align}
     \Pr\Set{\exists i\in A, \RK\neq `f_i(\RZ_i^n)} <`d \implies \Pr\Set{\RK\neq `q(\RG^n)} < `e\label{eq:KG}
 \end{align}
 for some functions $`f_A$ and $`q$.
\end{Lemma}

The recoverability constraint~\eqref{eq:recover} without private randomization implies the antecedent of the implication~\eqref{eq:KG} for sufficiently large $n$, and so the consequence of \eqref{eq:KG} holds with $`e$ going to $0$ sufficiently slowly in $n$, i.e.,
\begin{align*}
    \lim_{n\to `8} \Pr\Set{\RK\neq `q(\RG^n)} = 0.
\end{align*}
By Fano's inequality, this implies the property \eqref{eq:K|G}, i.e., the randomness of $\RK$ primarily comes from that of $\RG$. As in the converse proof in Section~\ref{sec:proof:CS0:untrusted} for the case with untrusted helpers, such property implies the desired upper bound $H(\RG|\RZ_D)$ on the secret key rate. 

\begin{Proof}[Proof of Lemma~\ref{lem:KG}]
 Similar to the proof of Theorem~\ref{thm:fls}, we will prove the Lemma by induction on $\abs{A}$. The base case follows from \cite[Lemma~1.1]{csiszar00}.\footnote{We set $\RX$ and $\RY$ in \cite[Lemma~1.1]{csiszar00} to $\RZ_1$ and $\RZ_2$ respectively.} 

 To prove \eqref{eq:KG} for general $A$, consider any $`e>0$ and $j\in A$, and let $\RZ'_j$ be the m.c.f.\ of $\RZ_{A`/\Set{j}}$.  Assume as an inductive hypothesis that, for any $`e'>0$, there exists $`d'>0$ independent of $n$ with
 \begin{align}
     \Pr\Set{\exists i\in A`/\Set{j}, \RK\neq `f_i(\RZ_i^n)} <`d' \implies \Pr\Set{\RK\neq `f'_j({\RZ'_j}^n)} < `e'\label{eq:KG:IH}
 \end{align}
 for some function $`f'_j$. Now, since $\RG$ is the m.c.f.\ of $\RZ_j$ and $\RZ_{A`/\Set{j}}$, we have again by \cite[Lemma~1.1]{csiszar00} that there exists $`d''>0$ independent of $n$ such that
  \begin{align*}
     \Pr\Set{\RK\neq `f_j(\RZ_j^n)\text{ or }\RK\neq `f'_j({\RZ'_j}^n)} <`d'' \implies \Pr\Set{\RK\neq `q(\RG^n)} < `e,
 \end{align*}
 which has the same consequence as the desired implication~\eqref{eq:KG}.
 By the union bound, the antecedent of the above implication holds if
 \begin{align}
    \Pr\Set{\RK\neq `f_j(\RZ_j^n)} &\leq `d''/2 \label{eq:KG:1}\\
    \Pr\Set{\RK\neq `f'_j({\RZ'_j}^n)} &\leq `d''/2.\label{eq:KG:2}
 \end{align}
 By the inductive hypothesis~\eqref{eq:KG:IH} with $`e'=`d''/2$, \eqref{eq:KG:2} holds if
 \begin{align}
     \Pr\Set{\exists i\in A`/\Set{j}, \RK\neq `f_i(\RZ_i^n)} <`d'\label{eq:KG:3}
 \end{align}
 for some $`d'>0$ independent of $n$. Choosing $`d=\min\Set{`d',`d''/2}$, which is independent of $n$, the antecedent of \eqref{eq:KG} implies both \eqref{eq:KG:1} and \eqref{eq:KG:3}, which imply  the consequence of \eqref{eq:KG} as desired.
\end{Proof}

We remark that the converse proof in Section~\ref{sec:proof:CS0:untrusted} is stronger because it applies to the more general case with public discussion at zero rate. If there is public discussion, the recoverability constraint~\eqref{eq:recover} may not imply the antecedent of \eqref{eq:KG} while Lemma~\ref{lem:KG} does not appear to extend to the case with public discussion. In particular, \cite[Lemma~1.1]{csiszar00} relies on a property of the Hirschfeld-Gebelein-R\'enyi maximal correlation that appears to fail to incoporate the public discussion.

\section*{Acknowledgment} 

The authors would like to thank Prof.\ Navin Kashyap for his valuable comments and contributions to the prior work in \cite{chan18isit}.

\bibliographystyle{IEEEtran}
\bibliography{IEEEabrv,ref}

\end{document}